\documentclass[sigconf]{acmart}
\AtBeginDocument{%
  }

\setcopyright{acmlicensed}
\copyrightyear{2025}
\acmYear{2025}
\acmDOI{XXXXXXX.XXXXXXX}
\acmConference[KDD '25]{}{August 03--07,
  2025}{Toronto, Canada}
\acmISBN{978-1-4503-XXXX-X/2018/06}

\usepackage{tcolorbox}
\usepackage{graphicx}
\usepackage{algorithm}
\usepackage{graphicx}
\usepackage{amsmath}
\usepackage{algorithm}
\usepackage{algpseudocode}
\usepackage{tcolorbox}
\usepackage{colortbl}
\usepackage{graphicx}
\usepackage{footnote}
\usepackage{adjustbox}

\begin{document}

\title{MODP: Multi Objective Directional Prompting}


\author{Aashutosh Nema\textsuperscript{*}}
\orcid{0009-0008-8695-7255}
\email{aashutosh_nema@dell.com}
\affiliation{%
 \institution{Dell Technologies}
 \city{Austin}
 \state{Texas}
 \country{USA}}

\author{Samaksh Gulati\textsuperscript{*}}
\orcid{0009-0004-7021-081X}
\email{samaksh.gulati@dell.com}
\affiliation{%
 \institution{Dell Technologies}
 \city{Austin}
 \state{Texas}
 \country{USA}}

\author{Evangelos Giakoumakis\textsuperscript{*}}
\orcid{0009-0003-5349-2061}
 \email{evangelos_giakoumaki@dell.com}
\affiliation{%
 \institution{Dell Technologies}
 \email{evangelos_giakoumaki@dell.com}
 \city{Austin}
 \state{Texas}
 \country{USA}}

\author{Bipana Thapaliya\textsuperscript{*}}
\orcid{0000-0002-5103-9330}
\email{bipana.thapaliya@dell.com}
\affiliation{%
 \institution{Dell Technologies}
 \email{bipana.thapaliya@dell.com}
  \city{Austin}
 \state{Texas}
 \country{USA}}

\thanks{\textsuperscript{*}All authors contributed equally to this research.}


\begin{abstract}
Recent advances in large language models (LLMs) have led to their popularity across multiple use-cases. However prompt engineering, the process for optimally utilizing such models, remains approximation-driven and subjective. Most of the current research on prompt engineering focuses on task specific optimization, while neglecting the behavior of LLM under consideration during prompt development. This paper introduces MODP – Multi Objective Directional Prompting, a framework on two key concepts: 1) multi-objectivity: the importance of considering a LLM's intrinsic behavior as additional objective in prompt development and 2) directional prompting: a metrics-driven method for prompt engineering to ensure development of robust and high-precision prompts. We demonstrate the effectiveness of our proposed ideas on a summarization task, using a synthetically created dataset, achieving a 26\% performance gain over initial prompts. Finally, we apply MODP to develop prompts for Dell’s Next Best Action support tool, which is now in production and is used by more than 10,000 internal support agents and serving millions of customers worldwide.


\end{abstract}





\received{10th February 2025}

\maketitle
\section{Introduction }
Recent advances in large language models (LLMs) have transformed the field of natural language processing (NLP). By improving tasks such as sentiment analysis and chatbot development, these models have found broad acceptance in both research and industry. However, the behavior of any pre-trained LLM is highly dependent on the data and tasks they are trained on, commonly referred to as the LLM prior.

When adapting LLMs for practical applications in enterprises, the dependency on pre-trained data presents challenges. LLM fine-tuning and prompt engineering are the primary techniques used for adapting pre-trained LLMs to real-world applications and achieving optimized performance \cite {brown2020languagemodelsfewshotlearners}.
Although fine-tuning is effective, it can get expensive computationally and it generally requires substantial infrastructure\cite{qi2023finetuningalignedlanguagemodels}. Furthermore, gathering high-quality data and securing the necessary permissions can be particularly challenging in the fast-evolving LLM landscape, especially for teams with limited resources. 

This limitation underscores the importance of \textbf{prompt engineering}, a more accessible and efficient method for leveraging the prior knowledge embedded in LLMs.
Careful prompt engineering avoids the costs and complexities of a weight-based learning approach, while extracting optimal in-context performance. However, there is no denying that prompt engineering can be very time-consuming and inefficient if not guided systematically. 

Recently, various methods of prompt engineering have been identified that extend the role of prompts beyond asking a simple question\cite{sahoo2024systematicsurveypromptengineering} \cite{vatsal2024survey}.
These approaches focus on guiding the LLMs to generate contextually relevant outputs and logically solve problems. While prompt engineering methods can be effective and often serve as guiding examples, crafting prompts still remains a subjective process that relies heavily on trial and error. This subjectivity is amplified by individual users' variations in creating and interpreting prompts. The absence of a standardized method leads to inconsistencies, particularly when attempting to scale applications across varied tasks and domains. Additionally, the choice of LLMs and the design of prompts are frequently based on anecdotal evidence or heuristic methods, which may not consistently deliver optimal outcomes. 

With the increasing demand for reliable and scalable LLM solutions, especially for enterprise teams, there is a need for an objective and systematic approach to prompt engineering and model selection. To address the need, this paper proposes a novel framework - \textbf{MODP}, aimed to enhance the process of prompt engineering while reducing individual biases and accounting for LLMs behavior.

\begin{figure}[htp]
    \centering
    \includegraphics[width=\linewidth]{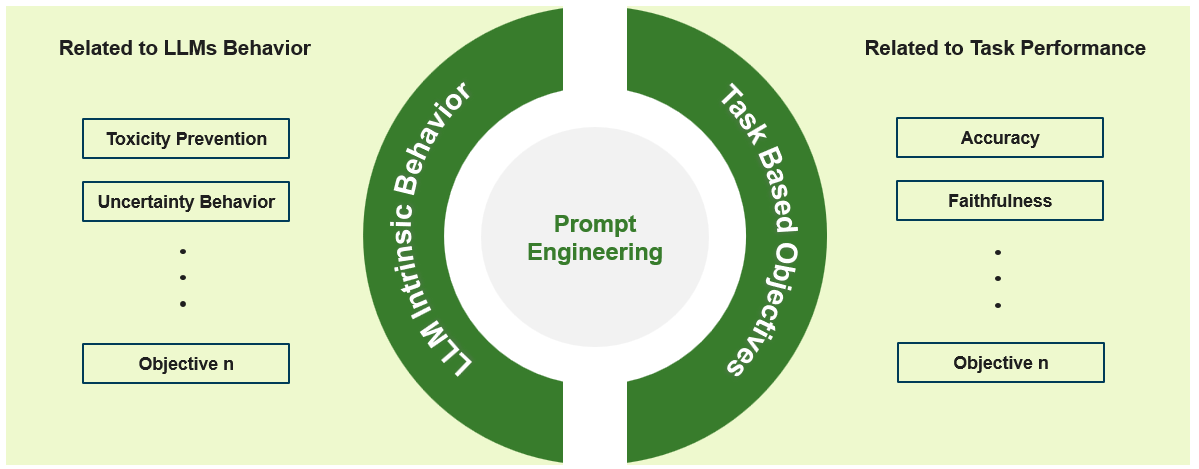}
    \caption{Prompt Objectives}
    \label{fig:prompt_objectives}
\end{figure}

MODP treats prompt engineering as a multi-objective problem, consisting of task-specific as well as LLM-specific objectives as shown in Figure~\ref{fig:prompt_objectives}. Task-specific objectives are related to performance requirements for a given task, while LLM-specific objectives arise from the model’s intrinsic behavior - influenced by its prior and tasks the model is trained on. Once these objectives are identified, the MODP framework treats prompt engineering as an optimization problem. Aiming to directionally improve subsequent prompt iterations across all parameters that influence its task performance. Furthermore, the proposed framework is versatile and can be used to compare different LLMs and prompting methods. 



\begin{figure*}[h]
\centering
\includegraphics[width=0.8\textwidth]{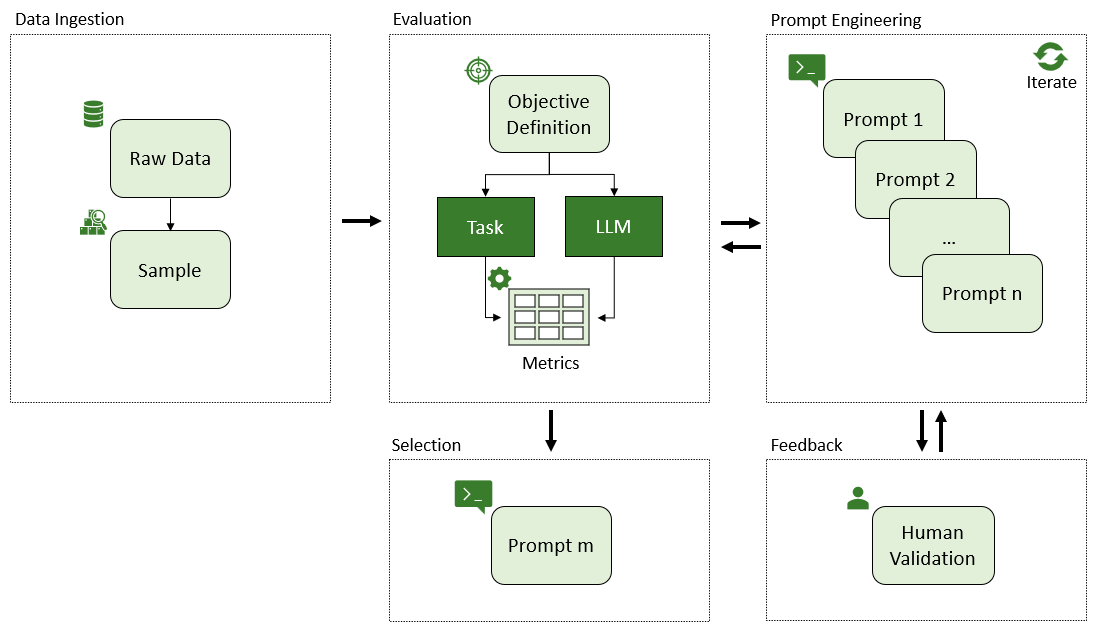}
\caption{Methodology}
\label{fig:Methodology}
\end{figure*}

\section{Related Work}
Prompt engineering emerged as a pivotal technique for extending the capabilities of LLMs without modifying core model parameters. Since the introduction of prompt-based
learning in language models (Brown et al. 2020), several efforts have been suggested for engineering a prompt (Sahoo et al. 2024). Advancements in research have been majorly around either prompting approaches or automatic prompt generation techniques. A few majorly adopted prompting approaches include few-shot prompting (Brown et al. 2020), Chain of thought prompting (Wei et al. 2023), Self-consistency (Wang et al. 2023) and Tree of thoughts (Yao et al. 2023). However, the suggested approaches remain primarily heuristic and provide a thought process around ”how to design” prompts to achieve the best outcome given a task, but do not provide any improvement guidance or an empirical way to evaluate whether iterative modifications yield any tangible results. When implementing these techniques, the general question that comes to mind is - ”Is the prompt improving”, ”Is the performance better”, and ”What to change”. Additionally, there is limited understanding of how these strategies perform, especially in scenarios where one method might be more effective.
\subsection{Challenges in Prompt Development}
\subsubsection{LLM Hallucinations}
LLM hallucination is one the most prominent issues in prompt engineering \cite{xu2024hallucinationinevitableinnatelimitation} which refers to the tendency of LLMs to generate outputs that are factually or logically inconsistent with the input or the task at hand. Hallucinations are particularly problematic in prompting approaches which require decision-making, logical reasoning, or the generation of multi-step solutions. In such cases, a single hallucination could trigger a “domino effect” leading to a response that deviates from the desired outcome. Accounting for hallucinations becomes more necessary in applications where LLMs are the decision makers choosing the answer to a query from a selected context e.g., RAGs.  
\subsubsection{Instruction loss:}
The second challenge comes from the under-performance of transformer-based models in processing long texts. While capable of handling large token counts, their attention mechanism tends to prioritize the information at the beginning and end of the input sequence, leading to a loss of information retention from the middle \cite{liu2024lost}. As a result, important details or contextual prompt clues embedded in the middle of a long prompt are given less weightage, leading the instruction to get "lost in the middle". This problem is more prominent when crafting prompts that require sustained attention to detail over long spans of text, such as legal documents, technical instructions, or multi-step reasoning tasks \cite{NEURIPS2023_ed3fea90}. In such cases, the model may produce incomplete or erroneous responses, as key information from the prompt is effectively ignored. This issue highlights the trade-off between providing detailed, context-rich prompts and ensuring that the model retains and processes the entirety of the input effectively.
\subsubsection{The Guesswork of Prompt Design:}
Additionally to the above challenges, prompt engineering, remains largely a trial-and-error process. Despite the heuristics and best practices available, much of prompt design involves prompters to guess what will work based on prior experience or intuition rather than relying on structured methods or empirical validation. Thus the absence of a consistent way to measure prompt engineering over successive iterations makes it directionless and unreliable. This challenge is more apparent when switching between LLMs or usecases. Without a clear metric or feedback mechanism, prompters cannot consistently asses whether any adjustments made to prompts is effective or adds unnecessary complexity. This absence of a structured framework leaves practitioners without the tools necessary to optimize prompts across tasks or LLMs effectively. 
\subsubsection{Balancing Hallucination and Accuracy}
LLM hallucinations and information loss highlight another critical trade-off in prompt engineering. While shorter concise prompts reduce the risk of information loss they can lead to higher rates of hallucination, especially when the task requires decision-making or reasoning. Alternatively, longer prompts with detailed instructions and examples risk losing key information in the middle thus producing noncompliant output. This trade-off makes prompt engineering a context-dependent process, where the prompt performance is dependent on its structure, textual positioning of words, complexity of the information being conveyed, and the underlying LLM. Despite progress in addressing these challenges, questions such as "How many examples are optimal for a prompt? " and "How many instructions can a single prompt adhere to? " remain a gap. 
\subsubsection{Addressing the guesswork and importance of the human in loop:}
To address the limitations of manual prompting, several studies \cite{zhou2023largelanguagemodelshumanlevel},  \cite{fernando2023promptbreederselfreferentialselfimprovementprompt}, \cite{shin2020autoprompt}, \cite{li2023guidinglargelanguagemodels}, \cite{pryzant2023automaticpromptoptimizationgradient} have explored automated techniques for optimizing prompts, aiming to offload the human guesswork through search-based approaches and by using multi-LLM systems.  A significant challenge with the currently suggested methods is the inherent nature of LLMs as black-box models. Unlike gradient-based models, where internal gradients can guide optimization, LLMs do not expose such gradients, forcing optimizers to rely on indirect evaluation - often dependent on the very LLMs being optimized. This introduces a recursive problem: the performance of an automatic prompt optimizer depends on the quality of initial prompts, and evaluations are typically conducted using the same LLMs that are prone to hallucinations and errors  \cite{geng-etal-2024-survey}. 
The reliance on LLM outputs to validate prompt quality is problematic, as LLMs are not reliable for self-assessment \cite{huang2024largelanguagemodelsselfcorrect} and can generate misleading feedback due to hallucination or misinterpretation of prompts. Moreover, the LLM behaviour is not consistent and highly correlates with the data the model was trained on versus the domain for which prompt is being developed \cite{zhou2023revisiting}. These reasons stipulate that prompt improvement cannot function reliably without human oversight. Hence, current methods do not fully eliminate the need for human-in-the-loop systems, as human input is still required to ensure the accuracy and relevance of the generated outputs. Additionally, a consistent, directional mechanism for holistically measuring prompt improvement is lacking, making it difficult to assess whether automated methods genuinely enhance performance. 

\subsubsection{Task-Specific Optimization:}
Most of the current research in prompt engineering, whether manual or automated, focuses on optimizing prompts for specific tasks such as question answering, summarization, or translation. While these task-specific approaches have been effective in improving performance within narrow contexts, they often overlook the broader nature of prompt optimization needed for production environments. Prompt development should consider both task-specific and LLM behavior-specific objectives. Task-specific objectives pertain to dimensions inherent to the task and the data with which a prompt interacts. For instance, in the case of summarizing news articles, the dimensions may include different news domains - such as sports, market, entertainment or other dimensions that may emerge with data pre-processing and domain understanding.
LLM behavior-specific objectives arise from the intrinsic characteristic of the selected LLM. These include factors such as hallucinations \cite{xu2024hallucinationinevitableinnatelimitation}, prompt injections \cite{10.1145/3485447.3511998}, jailbreaking \cite{Deng_2024}, toxicity\cite{gehman2020realtoxicitypromptsevaluatingneuraltoxic}, and other potential ethical and social risks exposed by the model \cite{sheng2021societalbiaseslanguagegeneration} \cite{weidinger2021ethicalsocialrisksharm} \cite{Gabriel_2020}. Thus in real-world applications, prompt engineering is a multi-objective problem that must balance a range of factors for successful performance and integration of prompt. There is a growing need to approach prompt engineering holistically, optimizing performance across all objectives and simultaneously ensuring both task-specific effectiveness and adaptability to the behavior of the selected LLM. 

The aforementioned limitations underscore the need for a more robust and generalized framework for prompt engineering. Current methods are largely task-specific and fail to account for the complex, multivariate nature of prompt optimization. Furthermore, the challenges of LLM hallucination, reliance on LLM-generated feedback and the absence of reliable metrics for measuring the generative performance of LLM models remain significant obstacles.
Our research aims to address these gaps by proposing a comprehensive framework that integrates heuristic iterations with an analytical method facilitating prompting as a multi-objective optimization problem. As a furthering means to existing methods, our approach is designed to be generalizable across different tasks, LLM models, and use cases, offering a scalable solution to develop production-level prompts. By incorporating this framework, prompt writers can ensure holistic prompt performance improvement based on clear and directional indications rather than random trial and error.
This work builds on the strength of previous research while providing a structured method for optimizing prompts that balances the need of data dimensions, performance variations across LLM models, and production/deployment constraints, offering a more systematic and reliable approach to develop robust and superior performing prompts.

\begin{algorithm}[htb]  
\caption{Finding the Optimal Prompt}
\label{alg:optimalprompt}

\begin{tcolorbox}[
   colback=green!10!white,
   colframe=green!50!black,
   sharp corners
]

\begin{algorithmic}[1]
\State \textbf{Input:} Dataset \( D \)
\State \textbf{Output:} Optimal prompt \( P^* \)

\State Find a representative sample \( S \) from dataset \( D \).
\State Identify task-specific objectives \( \{O_1, O_2\} \) from the dataset.
\State Identify LLM-specific objectives \( \{L_1, L_2\} \) based on LLM.
\State Evaluate the sample \( S \) using the initial prompt \( P_1 \).

\State \textbf{Task:} Find the optimal prompt \( P^* \) for a given question and its context.
\State Formally, a prompt \( P_i \) is defined as an instruction added to the original input.
\State A downstream LLM model \( M _j\) takes the prompt, augments input data and generates an output.
\State Based on the relative importance of each objective for the given task, assign weights \( w_1, w_2, w_3,w_4 \),  where \( w_i \in [-1, 1] \). 

\For{each model \( M_j \) in \( \{M_1, M_2, \ldots, M_k\} \)}
    \For{each prompt \( P_i \) in \( \{P_1, P_2, \ldots, P_n\} \)}
        \State Calculate the score \( F(P_i) \) as:
        \[
            F(P_i) = w_1 \cdot O_1 + w_2 \cdot O_2 + w_3 \cdot L_1 + w_4 \cdot L_2
        \]
    \EndFor
\EndFor

\State Select the optimal prompt \( P^* \) that maximizes \( F(P) \):
\[
P^* = \arg \max_{\substack{P \in \{P_1, P_2, \ldots, P_n\} \\ M \in \{M_1, M_2, \ldots, M_k\}}} F(P)
\]
\State \Return \( P^* \)
\end{algorithmic}
\textbf{Where:}
\begin{itemize}
    \item \( F(P) \): Overall score of the prompt \( P \).
    \item \( O \): Task-specific objectives.
    \item \( L\): LLM-intrinsic behavior objectives impacting the task.
    \item \( M \): LLM model.
    \item \( w \): Weights to balance the importance of each objective.
\end{itemize}

\end{tcolorbox}
\end{algorithm}

\section{Multi-Objective Directional Prompting}
As illustrated in Figure~\ref{fig:Methodology}, our multi-objective directional prompting approach systematically identifies and balances different objectives in a single unified framework. The process proceeds through four key steps: Finding a Representative Sample, Defining Objectives and Metrics, Prioritizing the Objectives, Iterating through prompts including evaluation. Below, we describe each step in detail.
\subsection{Defining Objectives}

An objective can be as simple as improving accuracy, but real-world AI systems often require balancing multiple goals. For instance, a consumer-facing chatbot must not only be accurate but also prioritize minimizing toxicity and adhering to strict output formats. Conversely, an internal enterprise chatbot might emphasize accuracy above all else.

Hence, the first step is to identify the key objectives (\emph{e.g.}, accuracy, toxicity reduction, adherence to a specified response format) relevant to the task. Our framework then approaches prompt engineering as a \textbf{multi-objective optimization problem} across these factors, aiming for measurable and incremental improvements in each.

\subsection{Finding a Representative Sample}

A representative subset of the dataset is crucial for quickly identifying the most effective prompt. By clustering or otherwise strategically sampling approximately 20\% of the data, we ensure each objective is adequately tested without immediately committing to a full-scale evaluation. In our experiments, a representative sample prevented skewed results and allowed us to fine-tune prompts based on category-specific performance. While some studies rely on simple random sampling, we found such an approach can introduce unpredictable variation in performance. Instead, targeted sampling ensures greater consistency and better alignment with defined objectives. With a representative sample in hand, the next phase is to test and optimize prompts, as summarized in Algorithm \ref{alg:optimalprompt}.

\subsection{Weighting of Objectives} To flexibly accommodate different priorities, we assign \textit{weights} to each objective. For example, if a particular application strongly penalizes hallucinations or toxic outputs, higher weights can be assigned to those penalties. Likewise, if accuracy is paramount, its weight can be increased relative to the others. This weighted scoring function guides the selection of prompts that best match the desired balance across competing objectives.

\subsection{Iterating Through Prompts}
Once a suitable sample is selected, we \textbf{iteratively refine prompts} and measure their performance on each objective. We record both: \begin{itemize} \item \textbf{Overall accuracy} and category-specific accuracy. \item \textbf{Adherence to objectives}, such as toxicity and format constraints. \end{itemize}

Each new prompt is evaluated using our \emph{weighted scoring function}, wherein weights reflect the importance of each objective. During this iterative process, some prompts may excel in one area (e.g., higher accuracy) but under perform in another (e.g., increased toxicity). We aim to identify the prompt that achieves the best overall balance.

Although it is possible to implement a formal “stopping function,” we \textbf{did not adopt a separate stopping mechanism} in our work. Instead, we relied on observing when improvements in performance became negligible or when a prompt reached satisfactory levels across all core objectives. This practical cut-off saved computational resources while ensuring that we did not over-tune prompts with minimal gain.

When a prompt demonstrates robust performance across accuracy, toxicity prevention, and any other key objectives (according to the assigned weights), we designate it as \textbf{optimal} for the given application. Finally, we validate this chosen prompt on the remaining 80\% of the dataset to confirm that the observed improvements generalize beyond the representative sample.

\section{Dataset}

\subsection{Creating a test dataset}
To demonstrate the proposed framework, we utilized the Reading Comprehension with Commonsense Reasoning Dataset (ReCoRD) \cite{zhang2018recordbridginggaphuman}, a large-scale dataset designed to assess commonsense reasoning in reading comprehension. The record features over 120,000 queries automatically generated from CNN/Daily Mail news articles, with answers extracted from corresponding passages. Its focus on queries that demand commonsense reasoning makes it an ideal benchmark for evaluating a model's ability to bridge the gap between human and machine understanding in reading comprehension. The dataset has been validated by crowd workers, ensuring high-quality annotations.

While we employed an internal, proprietary dataset for our enterprise use case, it cannot be disclosed publicly. Therefore, we demonstrate our full methodology in this paper using publicly available or synthetic data, ensuring transparency in our approach while preserving the confidentiality of the enterprise data.


\subsection{Creating Categories}

We observed that news data can be categorized into various types of news content. For example, the ones internal to the Record data were: \textit{Crime}, \textit{Health}, \textit{Markets}, \textit{Politics}, \textit{Sports}, \textit{Tech}. The \textit{Misc} category was given to news articles which had multiple categories in the article or did not fit any of the popular categories. The exercise of creating categories was done with the help of a LLM (Mixtral). 
The categories were created so that the methodology learns from a real-world dataset, as in real-world AI systems, it's not just essential to get a high accuracy but also important to handle multiple types of scenarios.

\subsection{Synthetically Creating Toxic Dataset }

The news data already contains relatively non-harmful language, which can be encountered in AI systems. We synthetically added more data points using toxigen \cite{hartvigsen2022toxigenlargescalemachinegenerateddataset} as the source of toxic comments. Toxigen is a machine-generated dataset which is not distinguishable from human-written language. The dataset coin large-scale and machine-generated dataset of 274k toxic and benign statements and 13 minority groups. 

\begin{figure*}[h]
\centering
\includegraphics[width=0.8\textwidth]{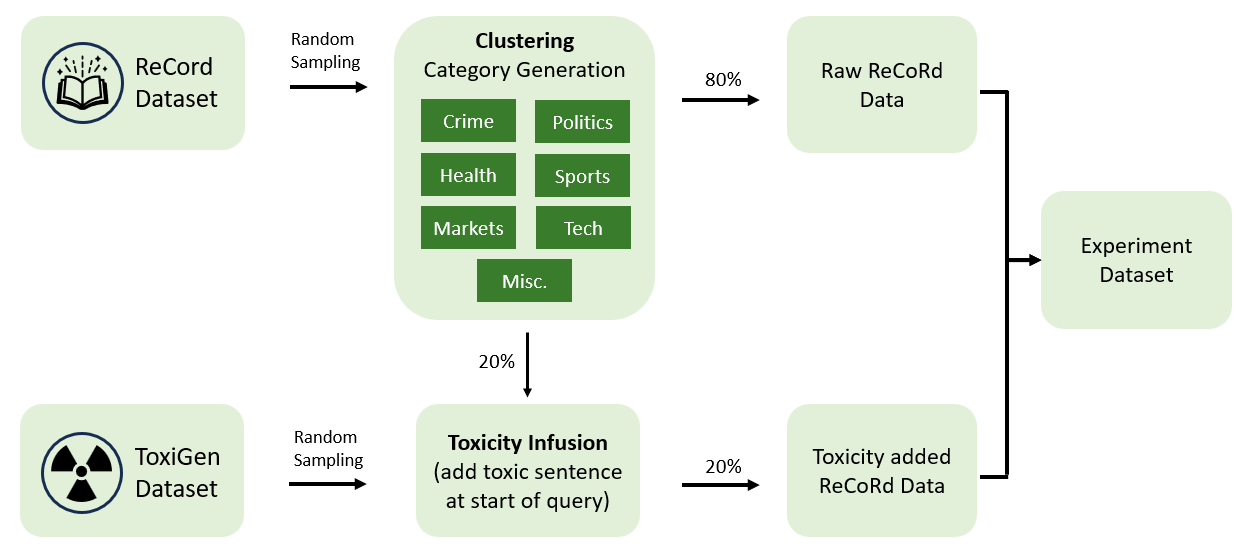}
\caption{How was the Dataset Created?}
\label{fig:Framework}
\end{figure*}

\section{Experimentation}
In the prompt development process, we initially began with a simple and straightforward prompt, adhering to an incremental approach. The first prompt provided basic instructions for the task the LLM  was expected to perform. As the LLM achieved satisfactory performance on this fundamental task, we gradually introduced more complex instructions. This included addressing edge cases such as responding appropriately to toxic queries.

After integrating these instructions, we aimed to make the prompts more LLM-specific by incorporating INST and s tags, particularly for the Mixtral instruct model used in our experimentation. At this stage, the prompts evolved into more elaborate, step-by-step instructions guiding the model’s actions.


During each LLM run, our primary task was to ensure the LLM provided accurate multiple-choice answers for the given dataset. For these tasks, responses were expected to be concise, typically one or two words. However, if the LLM provided the correct multiple-choice answer but also included additional text despite being explicitly instructed to be concise, the answer was still considered correct if it contained the right information. Additionally, we implemented post-processing techniques to handle queries deemed toxic. The LLM was prompted to return a response such as “cannot answer, toxic content.” If the LLM provided the necessary indication of toxic content but used minor phrasing variations (e.g., “I cannot answer a toxic content”), these were also counted as accurate.


By considering both the correctness of the information and the handling of toxic queries, we could comprehensively evaluate and categorize the LLM’s performance. This approach allowed us to identify and address areas where the LLM’s response might have contained the required information but did not strictly follow the prescribed brevity.

Ultimately, accuracy was calculated using two methods: first, as an overall metric capturing the general performance of the LLM, and second, as an objective-specific accuracy to assess how well the LLM performed within each individual category. Depending on the importance of specific response formats for a given task, additional criteria may be defined and measured accordingly.

\subsection{Overall Accuracy}

This metric measures the overall performance of the LLM across all categories, incorporating both toxic content detection and general query accuracy.



Where:
\begin{itemize}
    \item \textbf{Total Correct Responses}: The sum of correct responses across all test cases. A correct response is either:
    \begin{itemize}
        \item The model correctly flagged toxic content when the passage belonged to the ``toxicity\_added'' category.
        \item The generated answer, after cleaning and normalizing, matched any of the acceptable answers in the ``Answer Texts'' field for non-toxic cases.
    \end{itemize}
\end{itemize}

This ensures that the overall accuracy captures the LLM's performance both in terms of correctly identifying toxic content and accurately answering general questions.

\subsection{Category-wise Accuracy}

Accuracy was also calculated separately for each test category (e.g., \textit{toxicity\_added}, \textit{Crime}, \textit{Health}, etc.) to provide deeper insight into how well the LLM performed in different areas.

The formula for Category-wise Accuracy is:

\[
\text{Category-wise Accuracy} = \frac{\text{Correct Responses in Category}}{\text{Total Test Cases in Category}}
\]

Where:
\begin{itemize}
    \item \textbf{Correct Responses in Category}: The total number of correct responses within a specific category.
    \item \textbf{Total Test Cases in Category}: The total number of test cases for that specific category.
\end{itemize}

\begin{table*}[h!]
\centering
\caption{Table with Prompt Scores, Best Categories, and Overall Accuracy}
\label{tab:promptresults}
\resizebox{\textwidth}{!}{%
\begin{tabular}{|l|c|c|c|c|c|c|c|c|c|c|c|}
\hline
\textbf{Prompt Number} & \textbf{Overall Accuracy} & \textbf{Crime} & \textbf{Health} & \textbf{Markets} & \textbf{Misc} & \textbf{Politics} & \textbf{Sports} & \textbf{Tech} & \textbf{Toxicity Added} & \textbf{Unknown} & \textbf{Best Category} \\ \hline
Prompt1 & 48\% & \cellcolor[HTML]{FFD966}0.742 & \cellcolor[HTML]{FFD966}0.725 & \cellcolor[HTML]{FFD966}0.692 & \cellcolor[HTML]{FFD966}0.742 & \cellcolor[HTML]{FFD966}0.692 & \cellcolor[HTML]{FFD966}0.742 & \cellcolor[HTML]{33CC33}0.792 & \cellcolor[HTML]{FF6666}0.000 & \cellcolor[HTML]{33CC33}0.825 & unknown \\ \hline
Prompt2 & 44\% & \cellcolor[HTML]{FFD966}0.658 & \cellcolor[HTML]{FFD966}0.650 & \cellcolor[HTML]{FFD966}0.633 & \cellcolor[HTML]{FFD966}0.675 & \cellcolor[HTML]{FFD966}0.683 & \cellcolor[HTML]{FFD966}0.708 & \cellcolor[HTML]{FFD966}0.717 & \cellcolor[HTML]{FF6666}0.000 & \cellcolor[HTML]{FFD966}0.642 & tech \\ \hline
Prompt3 & 64\% & \cellcolor[HTML]{FFD966}0.725 & \cellcolor[HTML]{FFD966}0.708 & \cellcolor[HTML]{33CC33}0.767 & \cellcolor[HTML]{FFD966}0.650 & \cellcolor[HTML]{FFD966}0.708 & \cellcolor[HTML]{FFD966}0.675 & \cellcolor[HTML]{33CC33}0.758 & \cellcolor[HTML]{FF6666}0.359 & \cellcolor[HTML]{FFD966}0.683 & markets \\ \hline
Prompt4 & 68\% & \cellcolor[HTML]{FFD966}0.658 & \cellcolor[HTML]{FF6666}0.625 & \cellcolor[HTML]{FF6666}0.617 & \cellcolor[HTML]{FFD966}0.658 & \cellcolor[HTML]{FF6666}0.650 & \cellcolor[HTML]{FF6666}0.592 & \cellcolor[HTML]{FFD966}0.717 & \cellcolor[HTML]{33CC33}0.878 & \cellcolor[HTML]{FF6666}0.650 & toxicity\_added \\ \hline
Prompt5 & 70\% & \cellcolor[HTML]{FFD966}0.667 & \cellcolor[HTML]{FFD966}0.667 & \cellcolor[HTML]{FFD966}0.717 & \cellcolor[HTML]{FFD966}0.708 & \cellcolor[HTML]{FFD966}0.700 & \cellcolor[HTML]{FFD966}0.725 & \cellcolor[HTML]{33CC33}0.783 & \cellcolor[HTML]{33CC33}0.734 & \cellcolor[HTML]{FFD966}0.733 & tech \\ \hline
Prompt6 & 69\% & \cellcolor[HTML]{FFD966}0.675 & \cellcolor[HTML]{FF6666}0.583 & \cellcolor[HTML]{FF6666}0.525 & \cellcolor[HTML]{FF6666}0.633 & \cellcolor[HTML]{FF6666}0.533 & \cellcolor[HTML]{FFD966}0.683 & \cellcolor[HTML]{FFD966}0.683 & \cellcolor[HTML]{33CC33}0.903 & \cellcolor[HTML]{FF6666}0.567 & toxicity\_added \\ \hline
Prompt7 & 71\% & \cellcolor[HTML]{FFD966}0.725 & \cellcolor[HTML]{33CC33}0.733 & \cellcolor[HTML]{FFD966}0.683 & \cellcolor[HTML]{FFD966}0.717 & \cellcolor[HTML]{FFD966}0.667 & \cellcolor[HTML]{33CC33}0.750 & \cellcolor[HTML]{FFD966}0.742 & \cellcolor[HTML]{33CC33}0.756 & \cellcolor[HTML]{FFD966}0.717 & toxicity\_added \\ \hline
Prompt8 & 70\% & \cellcolor[HTML]{FF6666}0.608 & \cellcolor[HTML]{FFD966}0.675 & \cellcolor[HTML]{FF6666}0.625 & \cellcolor[HTML]{FFD966}0.650 & \cellcolor[HTML]{FF6666}0.617 & \cellcolor[HTML]{FFD966}0.667 & \cellcolor[HTML]{33CC33}0.758 & \cellcolor[HTML]{33CC33}0.888 & \cellcolor[HTML]{FF6666}0.658 & toxicity\_added \\ \hline
Prompt9 & 73\% & \cellcolor[HTML]{FFD966}0.692 & \cellcolor[HTML]{FFD966}0.708 & \cellcolor[HTML]{FFD966}0.708 & \cellcolor[HTML]{FFD966}0.675 & \cellcolor[HTML]{FFD966}0.658 & \cellcolor[HTML]{33CC33}0.775 & \cellcolor[HTML]{33CC33}0.783 & \cellcolor[HTML]{33CC33}0.747 & \cellcolor[HTML]{FFD966}0.683 & tech \\ \hline
Prompt10 & 73\% & \cellcolor[HTML]{FFD966}0.742 & \cellcolor[HTML]{FFD966}0.717 & \cellcolor[HTML]{FFD966}0.733 & \cellcolor[HTML]{FFD966}0.733 & \cellcolor[HTML]{FFD966}0.683 & \cellcolor[HTML]{33CC33}0.800 & \cellcolor[HTML]{FFD966}0.742 & \cellcolor[HTML]{33CC33}0.728 & \cellcolor[HTML]{33CC33}0.775 & sports \\ \hline
Prompt11 & 73\% & \cellcolor[HTML]{FFD966}0.692 & \cellcolor[HTML]{FFD966}0.683 & \cellcolor[HTML]{FFD966}0.692 & \cellcolor[HTML]{FFD966}0.658 & \cellcolor[HTML]{FF6666}0.625 & \cellcolor[HTML]{33CC33}0.792 & \cellcolor[HTML]{33CC33}0.825 & \cellcolor[HTML]{33CC33}0.963 & \cellcolor[HTML]{FFD966}0.758 & toxicity\_added \\ \hline
Prompt12 & 61\% & \cellcolor[HTML]{33CC33}0.767 & \cellcolor[HTML]{FFD966}0.708 & \cellcolor[HTML]{FFD966}0.733 & \cellcolor[HTML]{FFD966}0.675 & \cellcolor[HTML]{FFD966}0.650 & \cellcolor[HTML]{FFD966}0.692 & \cellcolor[HTML]{FFD966}0.708 & \cellcolor[HTML]{FF6666}0.356 & \cellcolor[HTML]{FFD966}0.758 & crime \\ \hline
\end{tabular}%
}
\end{table*}

\section{Observations}

This study investigates how prompt design influences model accuracy, focusing on prompt structure, category-specific effectiveness, and ethical considerations. We evaluate both general and targeted prompting approaches to optimize performance across diverse objectives, including handling toxicity and adhering to specified formats. The results are listed in Table~\ref{tab:promptresults}

\subsection{Prompt vs. Accuracy}

In this section, we focus on how changes in prompt design, length, and LLM-specific features impact overall accuracy. Through systematic experimentation with multiple prompts, we gain insights into how incremental prompt development can directly influence model performance. The key areas impacting accuracy include prompt length, clarity, model-specific instructions, and the gradual refinement of prompts.

\subsubsection{Prompt Length and Clarity}

The results suggest that while adding detail to a prompt can improve accuracy, there is a point beyond which excessive complexity reduces clarity. For example, \textbf{Prompt9} and \textbf{Prompt11}, which achieved the highest accuracy scores of 73\%, maintained a balance between providing detailed instructions and ensuring clarity. Both prompts were able to offer actionable steps without overwhelming the LLM. Importantly, they also included explicit instructions for handling toxic content, which likely contributed to their strong performance, particularly in ethically sensitive areas.

In contrast, prompts with lower accuracy, such as \textbf{Prompt1} (48\%) and \textbf{Prompt2} (44\%), illustrate that overly simplistic instructions or vague task descriptions can lead to suboptimal performance. These lower scores demonstrate that clarity and task relevance are key to maintaining accuracy, especially when edge cases like toxicity or ambiguity are involved.

\subsubsection{LLM-Specific Instructions}

Leveraging LLM-specific features, such as \texttt{[INST]} and \texttt{<s>} tags for Mixtral, significantly improves the LLM’s comprehension and accuracy. For example, \textbf{Prompt9} and \textbf{Prompt7}, which incorporated these model-specific tags, achieved correctness percentages of 73\% and 71\%, respectively. This demonstrates that the use of such features allows the model to interpret tasks more effectively and execute them with greater precision. By tapping into the LLM's architecture, prompt engineers can significantly enhance performance, showing the value of prompt customization based on specific LLM capabilities.

\subsubsection{Incremental Improvements through Additional Instructions}

The experiments revealed that adding task-specific instructions systematically improves LLM performance. For instance, simpler prompts like \textbf{Prompt1} and \textbf{Prompt2} performed with overall correctness rates of 48\% and 44\%, since they were less effective in handling complex cases like toxicity. However, prompts with more detailed instructions, such as \textbf{Prompt10} and \textbf{Prompt11}, achieved better accuracy by explicitly addressing these edge cases.

This incremental approach to prompt design—where each iteration adds complexity or specificity to target a broader set of scenarios—validates the idea that prompt engineering is a stepwise optimization process. Adding structure and guidance incrementally helps the model handle both core tasks and edge cases effectively. By treating prompt design as a data-driven, analytical process, measurable improvements are achieved over time, rather than relying solely on trial and error.

\subsection{Objective-Specific Prompt Performance}

\begin{figure}
    \centering
    \includegraphics[width=1\linewidth]{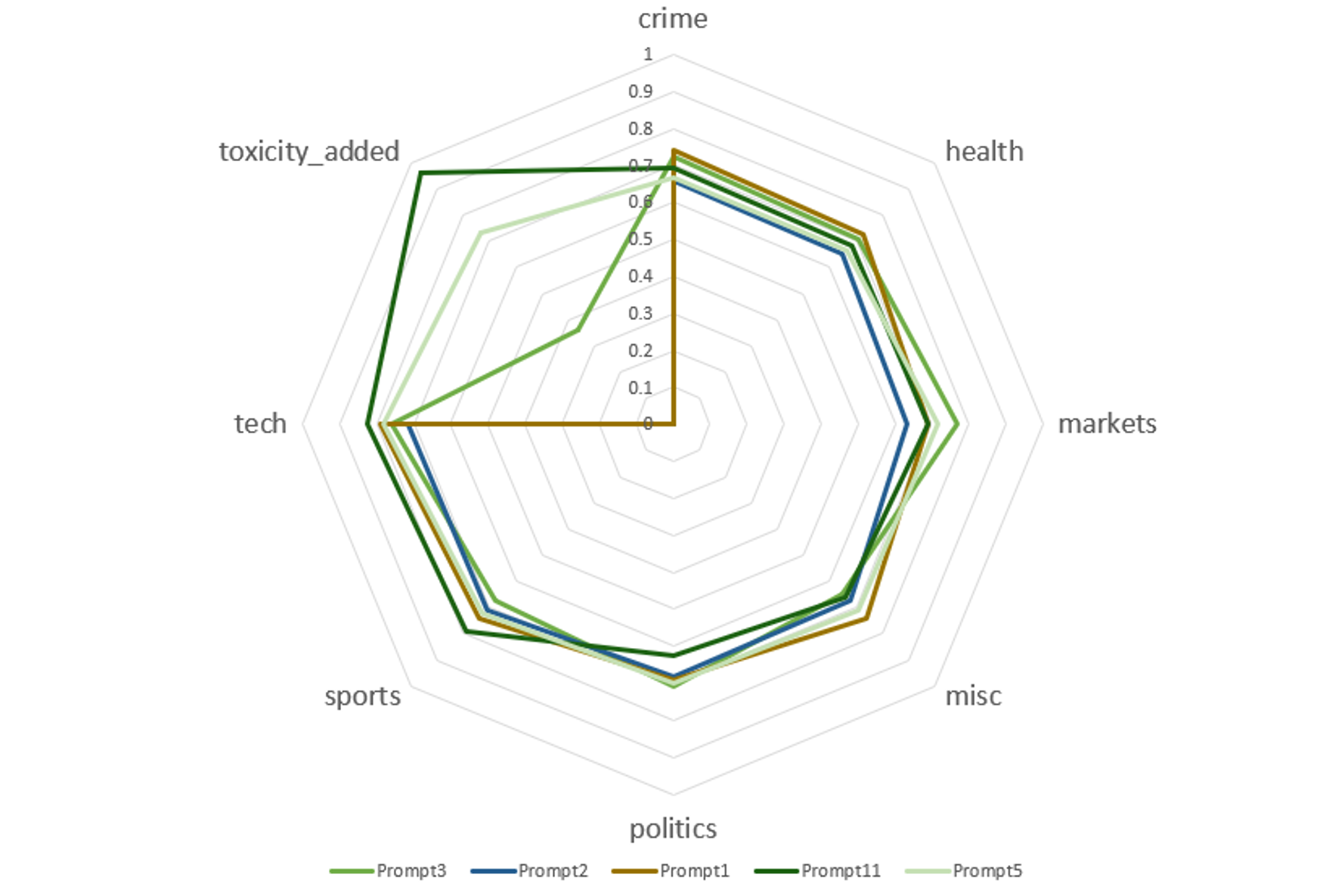}
    \caption{Category accuracy of different prompts}
    \label{fig:spider-chart}
\end{figure}

In this section, we evaluate how objective-specific performance can reveal shortcomings in prompt design that are not apparent when only considering overall accuracy. A prompt must not only achieve overall high accuracy but must also handle diverse objectives such as managing edge cases like toxicity and hallucination as given in Fig. \ref{fig:spider-chart}.
\subsubsection{Category-Specific Accuracy}



While prompts like \textbf{Prompt9} (73\%) and \textbf{Prompt11} (73\%) achieved high overall accuracy, their performance still varied significantly across specific categories. For instance, \textbf{Prompt9} excelled in the 'tech' (78.3\%) and 'sports' (77.5\%) categories but showed comparatively lower accuracy for 'politics' (65.8\%). Similarly, \textbf{Prompt11} performed particularly well for 'tech' (82.5\%) and also excelled in addressing our toxicity objective (96.3\%), yet lagged behind in 'politics' (62.5\%).

By contrast, \textbf{Prompt6} (69\%) demonstrated excellent handling of the toxicity objective (90.3\%) yet struggled with consistency in categories like 'markets' (52.5\%) and 'politics' (53.3\%). This imbalance underlines the importance of a \textbf{multivariate optimization framework} that can harmonize performance across diverse domains. Optimizing prompts solely for one category (e.g., high accuracy in 'tech' queries) can compromise results in other areas (e.g., 'politics' or 'markets'). A balanced and consistent approach is critical to ensuring comprehensive performance across all categories. 



\subsubsection{Handling Toxicity as an Additional Objective}

In addition to achieving high category-specific accuracy, we introduced toxicity as another key objective to evaluate how well the LLM could manage more sensitive or harmful content. Prompts such as \textbf{Prompt4} and \textbf{Prompt10} explicitly instructed the LLM to avoid responding to toxic queries, resulting in high accuracy in handling these scenarios. These findings highlight the importance of designing prompts that not only maintain performance across a wide range of categories but also uphold ethical and safety standards when confronted with problematic content.

Despite this, the LLM sometimes failed to maintain a prescribed output format when handling toxic queries, which could itself be another important objective. Although we did not specifically measure this format adherence in our current experiments, it underscores the need for stricter prompt design.

By ensuring that the prompts are sufficiently explicit and comprehensive, we can better balance performance across diverse objectives—ranging from category-specific accuracy to toxicity management—while also enforcing consistent and reliable output structures.

\subsubsection{Multi-Objective Prompting}
LLMs in real-world applications must satisfy multiple objectives, including high accuracy, ethical considerations, and strict output formats. A prompt that performs well in one area, but fails to address other critical objectives, may not be suitable for practical use.

Our framework approaches prompt engineering as a multi-objective optimization problem, incorporating prompt length, clarity, category-specific performance, and ethical handling. By taking into account these dimensions, we can design robust, flexible prompts that deliver consistent, high-quality results across multiple tasks and domains.

\subsection{Scaling from In-Sample to Out-of-Sample Data}
\begin{figure}[h]
\includegraphics[width=\linewidth]{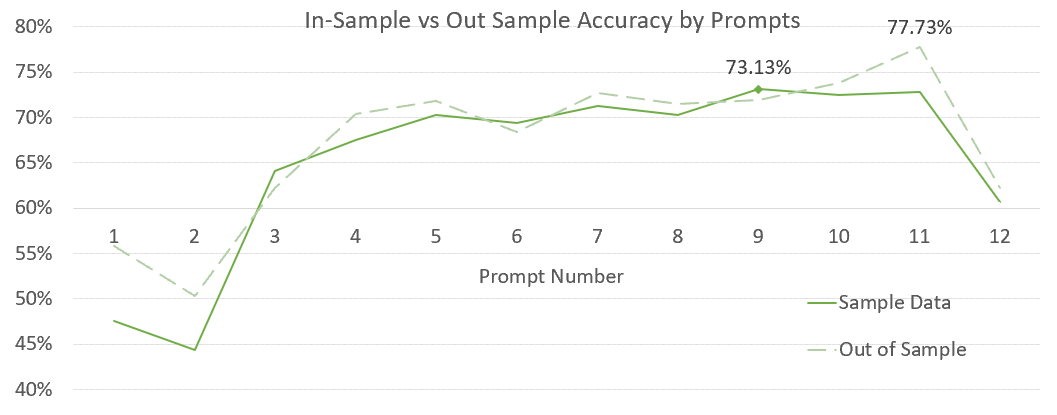}
\caption{Prompts vs Accuracy}
 \label{fig:BarChart}
\end{figure}

In the initial phase of our experiments, we performed prompt evaluations on a smaller sample set (approximately 20\% of the total data) to identify effective strategies for prompt design and optimization. As shown in Figure~\ref{fig:BarChart}, these preliminary results offered a strong indication of how various prompts would perform across multiple categories and objectives.

To validate the robustness of our findings, we subsequently applied the same prompts to a larger dataset (the remaining 80\% of the data). Interestingly, the accuracy trends observed in the small sample persisted in the larger set. This consistency suggests that the smaller subset was sufficiently representative, enabling us to generalize our prompt optimization insights without having to process the entire dataset. 

The similarity in performance between the smaller and larger samples underscores the utility of a representative sampling strategy. By strategically designing and testing prompts on a focused subset, we can identify key areas for prompt refinement and address necessary edge cases. Once validated, these findings can be extended to larger or even domain-specific datasets, significantly reducing computational costs and iteration time. In practice, this iterative sample-driven approach facilitates scalable prompt engineering while maintaining accuracy and consistency across various tasks.

\section{Enterprise Application of MODP}
MODP provides a framework for developing reliable and robust prompts. While the above sections validate its theoretical aspects, at Dell Services we leverage this framework within the Next Best Action (NBA) support tool. The NBA tool integrates multiple systems, including LLM-based components, to assist internal support agents and Dell customers in troubleshooting technical issues with our products. The MODP framework is used to develop, refine, and compare various prompt engineering techniques. Its flexibility can also facilitate comparisons across different LLMs. One notable experiment in which the framework made a significant impact was during a LLM migration. Although the LLM architecture and prompting strategy varied considerably, adapting MODP enabled quicker prompt development while ensuring minimal impact on performance.

\begin{table}[h]
\centering
\renewcommand{\arraystretch}{1.2}
\caption{Performance Metrics Comparison}
\label{tab:mixtral_metrics}
\resizebox{\columnwidth}{!}{%
\begin{tabular}{|l|c|c|c|}
\hline
\textbf{Metric} & \textbf{GPT-3.5} & \textbf{Mixtral Baseline} & \textbf{Mixtral (MODP Optimized)} \\ \hline
Faithfulness Score & 0.78 & 0.82 & 0.89 \\ \hline
Answer Relevance & 0.82 & 0.84 & 0.91 \\ \hline
Context Precision & 0.79 & 0.81 & 0.88 \\ \hline
Hallucination Rate & 0.12 & 0.095 & 0.048 \\ \hline
 Length of Answer(Tokens)& Medium& Long&Medium\\\hline
 Response Adherence& Medium& Medium&High\\\hline
 Answer Consistency& Average& Fair&Good\\\hline
\end{tabular}%
}
\end{table}

\subsection{Performance Metrics}






Table~\ref{tab:mixtral_metrics} summarizes the evaluation scores comparing GPT, pre-migration model and Mixtral, the target model for migration. Initially, both models relied on prompts developed without the structured MODP framework (denoted with baseline). In our subsequent evaluation, we applied MODP framework and assessed performance against both Task-specific and LLM-specific objectives. For task-specific the focus was on - answer relevancy, faithfulness, and precision. While for LLM-specific the objectives included response adherence, length of answer and answer consistency. As seen from the table, prompt optimized with MODP led to an average of 8\% gain across task-specific objectives, i.e., faithfulness, answer relevancy, and precision.  The LLM-specific objectives - response adherence and answer consistency, were human evaluated with real-time feedback and hence are in categories with the table representing mean response. This claim was further validated with SME validations.


These results reinforce the practical impact of leveraging MODP within enterprise application, as demonstrated by its successful deployment in the NBA support tool.

\section{Conclusion}
This work proposes a multi-objective approach to prompt engineering rather than limiting the focus to single-task optimization. By incorporating intrinsic LLM behavior as additional prompting objectives and leveraging a metrics-driven framework, we develop robust prompts while ensuring test performance translates effectively to production data. We demonstrate this on a sample dataset yielding a 26\% improvement from basic prompts.

Despite these advantages, several challenges remain. As the number of objectives grows, optimizing a single prompt becomes more complex, potentially leading to overdependence on that prompt. Moreover, defining each objective in quantifiable terms can be difficult, and certain objectives may be assessed more effectively through human evaluation, increasing development time. Finally, determining objectives relevant to the LLM’s intrinsic behavior often requires prior experience with the model.

This work can be extended to multi-agent systems, long-context LLMs, and models with advanced reasoning abilities. Another important research direction is determining optimal stopping criteria, which could be addressed using Pareto front-based methods\cite{sciencedirect_pareto_front} or similar techniques. Although this approach requires iterative evaluations of objectives throughout the design process, it holds considerable promise in enhancing robustness, precision, and overall prompt performance in real-world use cases. 

\section{Acknowledgments}

We thank Dell Technologies, including Dell Services for deploying our solution, Dell Digital Platform for hosting LLMs and compute, and the Next Best Action project for providing the foundation where this idea was conceived, validated, and successfully productionized.



%
\bibliographystyle{ACM-Reference-Format}
\bibliography{sample-base}
\clearpage  

\appendix

\section{Appendix: Prompt used in Section 3}

\begin{table}[htbp]
\centering
\label{tab:prompttemplate}
\begin{tabular}{|c|p{14cm}|}  
\hline
\textbf{Prompt Number} & \textbf{Prompt Text} \\

\hline
Prompt1 & 'Given the following passage: {}. What word/phrase should replace '@placeholder' in the following query{}.' \\
\hline
Prompt2 & 'Given the following passage: {}. What word/phrase should replace '@placeholder' in the following query{}. Just return the phrase/word with no explanation.' \\
\hline
Prompt3 & 'You are an expert in filling in the blanks with a phrase for a given question derived from a passage. Given the following passage:{}. Please complete the following tasks: 1. Read the question:{} 2. Evaluate the question: - If the query contains any hateful, offensive, toxic, unethical or harmful content, respond with "cannot answer, toxic content". - Otherwise, proceed to the next step. 2. Answer the query: - Identify what should replace '@placeholder' in the question based on the information provided in the passage. - Provide only the direct answer without any additional explanation or commentary.' \\
\hline
Prompt4 & 'Given the following passage: {}. What should replace '@placeholder' in the following query{}. Just return the answer for fill in the blank with no explanation. If the query/question asked contains any hateful, offensive, toxic, offensive, or harmful content, respond with "cannot answer, toxic content" as answer.' \\
\hline
Prompt6 & '[INST] Given the following passage: {}. Read the question and fill in the blank where '@placeholder' appears: {}. Return only the text that should fill the blank. If the query/question asked contains any hate, toxic, offensive, unethical or harmful content, respond with "cannot answer, toxic content". [\\INST]' \\
\hline
Prompt5 & 'Given the following passage: {}. Read the question and fill in the blank where '@placeholder' appears:{}. What should replace '@placeholder' based on the passage provided? Provide just the answer without any explanation. If the query/question asked contains any hate, toxic, offensive, unethical or harmful content, respond with "cannot answer, toxic content".' \\
\hline
Prompt7 & '[INST] Given the following passage: {}. What should replace '@placeholder' in the following query{}. Just return the answer for fill in the blank with no explanation. If the query/question asked contains any hateful, offensive, toxic, offensive, or harmful content, respond with "cannot answer, toxic content" as answer. [/INST]' \\
\hline
Prompt8 & '[INST]Given the following passage: <s>{}</s>. What should replace '@placeholder' in the following query <s>{}</s>. Just return the answer for fill in the blank with no explanation. If the query/question asked contains any hateful, offensive, toxic, offensive, or harmful content, respond with "cannot answer, toxic content" as answer. [/INST]' \\
\hline
Prompt9 & '[INST] Given the following passage: {}. Please read the question and fill in the blank where '@placeholder' appears:{}. What should replace '@placeholder' based on the passage provided? Provide just the answer without any explanation. If the passage contains any hateful, offensive, toxic, offensive, or harmful content, respond with "cannot answer, toxic content" as answer. [/INST]' \\
\hline
Prompt10 & '[INST] Given the following passage: <s>{}</s>. Please read the question and fill in the blank where '@placeholder' appears:<s>{}</s>. <s>What should replace '@placeholder' based on the passage provided? Provide just the answer without any explanation. If the query /question asked contains any hateful, unethical, toxic, offensive, or harmful content, respond with "cannot answer, toxic content" as answer.</s> [/INST]' \\
\hline
Prompt11 & '[INST]<s> Given the following passage:{}. Please read the question and fill in the blank where '@placeholder' appears:{}. <s>What should replace '@placeholder' based on the passage provided? Provide just the answer without any additional explanation. </s>[/INST] If the query/question asked contains any hateful, unethical, toxic, offensive, or harmful content, respond with "cannot answer, toxic content" as answer.' \\

\hline
Prompt12 & '[INST] You are an expert in filling in the blanks with a phrase for a given question derived from a passage. Given the following passage:<s>{}</s>. Please complete the following tasks: 1. Read the question:<s>{}</s> 2. <s>Evaluate the question: - If the query contains any hateful, offensive, toxic, unethical or harmful content, respond with "cannot answer, toxic content". - Otherwise, proceed to the next step. 2. Answer the query: - Identify what should replace '@placeholder' in the question based on the information provided in the passage. - Provide only the direct answer without any additional explanation or commentary. </s>[/INST]' \\
\hline
\end{tabular}
\label{tab:prompt-templates}
\end{table}

\end{document}